\documentstyle[twocolumn,aps,psfig]{revtex}
\begin{document}
\draft
\twocolumn[\hsize\textwidth\columnwidth\hsize\csname @twocolumnfalse\endcsname
%
%
%

\title{Influence of Nearest-Neighbor Coulomb Interactions \\
 on the Phase Diagram of the  Ferromagnetic Kondo Model}

\author{A. L. Malvezzi, S. Yunoki, and E. Dagotto}

\address{National High Magnetic Field Lab and Department of Physics,
Florida State University, Tallahassee, FL 32306}

\date{\today}
\maketitle

\begin{abstract}
The influence of a nearest-neighbor Coulomb repulsion of strength $V$ on the properties
of the Ferromagnetic Kondo model is analyzed using computational
techniques. The Hamiltonian studied here is defined on a chain
 using localized $S=1/2$ spins, and one orbital per site. Special
emphasis is given to the influence of the Coulomb repulsion on the regions of
{\it phase separation} recently discovered in this family of models, as
well as on the
double-exchange-induced ferromagnetic ground state. When phase
separation dominates at $V=0$, the Coulomb interaction 
breaks the large
domains of the two competing phases into small ``islands'' of one phase
embedded into the other. This is in agreement with several 
experimental results, as discussed in the text.
Vestiges of the original phase separation regime
are found in the spin structure-factor 
as incommensurate peaks, even at large values of $V$. In
the ferromagnetic regime close to density $n=0.5$, the Coulomb interaction
induces tendencies to charge ordering without altering the fully
polarized character of the state. This regime of ``charge-ordered
ferromagnetism'' may be related with experimental observations of a
similar phase by C. H. Chen and
S-W. Cheong (Phys. Rev. Lett. ${\bf 76}$, 4042 (1996)). Our results
reinforce the  recently introduction notion (see e.g., S. Yunoki et al.,
Phys. Rev. Lett. {\bf 80}, 845 (1998))
that in realistic models for manganites analyzed with unbiased many-body
techniques, the ground state properties arise from a competition
between ferromagnetism and phase-separation/charge-ordering tendencies.

\end{abstract}
\vskip2pc]
\narrowtext

%
%

\section{Introduction}

The study of manganite materials continues attracting considerable 
attention due to their potential technological applications and
interesting physical properties dominated by its colossal
magnetoresistance effect~\cite{jin}. Ferromagnetism in models for
manganites has been studied since the 1950s, and it is widely assumed
that the Double-Exchange (DE) ideas are sufficient to understand the
stabilization of a magnetic state at low-temperatures upon hole-doping 
the manganites~\cite{zener}. 
The main reason for this stabilization is 
that the kinetic energy of the holes is substantially
improved in a fully aligned spin background. These concepts are similar
to those that lead to 
the well-known Nagaoka phase in the context of the $t-J$ model 
for the cuprates. However,
experimental work has revealed a phase diagram for manganites that is
far more complicated than simple DE models predict~\cite{co}. 
In particular,
at densities corresponding to the
undoped compound (referred to as ``half-filling'' ($n=1$) in the language of the
one-orbital model)
an A-type antiferromagnet (AF) is stabilized. After a small
amount of holes is introduced, some manganites enter a regime where
ferromagnetic droplets have been recently observed~\cite{hennion}.
Upon further doping 
the DE ferromagnetic
metallic regime is reached, including the widely studied $x \sim 1/3$
concentration. 
In the other limit of intermediate- and small-electron densities, a
charge-ordered state is stabilized in $\rm Ca$-doped manganites~\cite{co}. 
In addition, at densities corresponding
to the DE ground state at low-temperatures, a metal-insulator transition occurs
as the temperature is raised. This insulating state is likely
responsible for the colossal magnetoresistance effects in
manganites.

Recent work has initiated the computational analysis of models for
these compounds~\cite{yunoki,dago,yunoki2,yunoki3}, allowing for calculations that go 
beyond simple mean-field approximations. The
techniques used for such computational studies are borrowed in part from the
field of cuprates, where they were applied to the analysis of $t-J$-like
models in recent years~\cite{review}. The use of unbiased techniques in models
for manganites
has already led to a fairly good understanding of the zero temperature phase
diagram of the so-called Ferromagnetic Kondo model (FKM) which
is based on the assumption that only one $e_g$-orbital is relevant.
Although  Jahn-Teller (JT) effects cannot be neglected in a
quantitative study of manganites~\cite{millis}, 
it is important to clarify the properties of this simplest one-orbital
model before addressing more complicated two-orbital Hamiltonians with 
JT-phonons. 

Following this computational approach, 
Yunoki et al.~\cite{yunoki,dago,yunoki2} recently reported the phase
diagram of the FKM addressing dimensions $D=1,2,3$
and $\infty$. 
The computational techniques used were the Monte Carlo
method in $D=1,2,3$ with classical $t_{2g}$-spins, the dynamical
mean-field approximation in $D=\infty$~\cite{furu},
as well as the Exact Diagonalization (ED) 
and Density-Matrix Renormalization Group (DMRG)~\cite{white}
techniques in one dimension for quantum-mechanical localized $t_{2g}$-spins. 
The phase diagram was found to contain three dominant
regimes: (i) a ferromagnetic phase induced by the DE mechanism, (ii) a spin
 incommensurate regime at intermediate and small Hund
coupling~\cite{inoue}
predominantly due to the RKKY interaction, and
(iii) a region of ``phase separation'' (PS) near density $n=1$ between
ferromagnetic hole-rich and antiferromagnetic hole-undoped phases. The
latter regime (PS) is unexpected since for a long time it has been assumed
that the transition from the AF regime, characteristic of the undoped
region, to the FM regime at $x \sim 1/3$ should occur through a spin-canted
state~\cite{gennes}. However, the results of
 Ref.~\cite{yunoki} showed that this is incorrect and
actually the transition from AF to FM occurs through
the creation of ``islands'' of one phase embedded into the other, growing in
size as the overall density changes. Work by other groups~\cite{also} is in agreement 
with the main conclusions of Ref.~\cite{yunoki}. Previous studies in 1D
models also indicated the relevance of phase separation
as one moves from copper to the left 
in the transition-metal row of the
periodic table of elements~\cite{riera}.

On the experimental side, a growing body of evidence suggests that the
transition from undoped $\rm LaMnO_3$ to the doped ferromagnetic compounds
indeed occurs through an inhomogeneous process. For instance, the existence of
quasi-static
magnetic droplets at small hole-density in $\rm La_{1-x} Ca_x Mn O_3$
has recently been emphasized
using elastic neutron scattering techniques at $x=0.05$ and $0.08$ below
the critical temperature for the magnetic state~\cite{hennion}.
Phase separation using NMR techniques has also been observed for the same
compound at low temperatures~\cite{allodi}. Working at $x=1/3$
and at temperatures above the ferromagnetic critical temperature $T_c^{FM}$,
neutron scattering
experiments were interpreted as providing evidence for two distinct
phases in the system~\cite{lynn}, while small-angle neutron scattering and magnetic
susceptibility data suggested the presence of  magnetic
clusters ($\sim 12 \AA$ in size) in the same regime 
~\cite{deteresa}. X-ray and powder neutron
scattering methods applied to $\rm Pr_{0.7} Ca_{0.3} Mn O_3$ also revealed
the existence of ferromagnetic clusters~\cite{cox}. Neutron studies of
$\rm La_{1-x} Sr_x Mn O_3$ at $x=0.10$ and $0.15$ were interpreted as
corresponding to a polaron-ordered arrangement~\cite{yamada}. 
Experimental studies of the 2D compound
 $\rm Sr_{2-x} La_x Mn O_4$ at small electronic density 
have also been
considered as evidence of phase separation in manganites~\cite{bao}.
Many other papers have reported results compatible with charge
segregation tendencies at temperatures and densities that surround
 the low-temperature
ferromagnetic phase of the manganites~\cite{otroexp,perring,otro}.

This interesting agreement between theory and experiments
reinforces the notion that simple electronic models for manganites with
a strong Hund-coupling may contain the essence of the physics needed to
describe these materials. However, to confirm this assumption 
it is important to proceed further with the
computational calculations in two main directions: (i) first, the
influence of Jahn-Teller phonons and the use of two $e_g$-orbitals is
important for a qualitative comparison theory-experiment. An effort in
this direction has been recently reported by two of the authors and
A. Moreo~\cite{yunoki3}. 
The existence of PS with JT-phonons was confirmed,
including a tendency to  a novel 
``orbital'' phase separation regime~\cite{yunoki3}. PS
tendencies were found to be as robust in the phase diagram as in the case of a
one-orbital Kondo model. This shows that studies of one-orbital
Hamiltonians can capture at least part of the physics of more sophisticated
multi-orbital models for manganites; (ii) second,  
Coulomb interactions beyond the on-site term are important to
avoid the accumulation of charge suggested by the 
phase separation process described in Ref.~\cite{yunoki}. It is likely
that in the region of ``unstable'' densities
the large clusters of the
hole-rich ferromagnetic phase will be divided into small clusters due to this
Coulomb repulsion. 
Actually previous work in the context of the cuprates has
shown the tendency to form extended charge-ordered structures~\cite{emery} 
once a long-range
Coulomb interaction is added to a phase separated state. 

It is precisely one of
the purposes of the present paper to study the influence of Coulomb
interactions beyond the on-site $U$-term on a model for manganites that
has the tendency to phase separate in its ground state. The selection
of the model for the present investigation is important for technical reasons.
The Ferromagnetic Kondo model with classical $t_{2g}$-spins analyzed in
Ref.~\cite{yunoki} is difficult to study with Coulomb interactions
since in this case the Monte Carlo method will need the addition of
Hubbard-Stratonovich (HS) degrees of freedom
 to decouple the four-fermion terms in the Hamiltonian.
In addition to
 the intrinsic complexity of this procedure, the HS decoupling will
lead to ``sign-problems''~\cite{review} that likely will prevent the
study of low-temperature properties. 
To complicate matters even more,
techniques that deal directly with the Hilbert space of the problem,
such as ED and DMRG, cannot be
easily applied to a model with classical spins. Then, in order to
analyze the influence of intersite Coulomb terms it would be
better to use $quantum$ $t_{2g}$-spins and ED or
DMRG methods for their analysis. In this context there are no
``sign-problems''. However, the huge Hilbert spaces that need to be
studied in finite size chains 
impose constraints on the value of the quantum localized spins $S$.
In order to carry out a study on a cluster large enough to reach bulk
properties $S$ must be restricted to $1/2$, rather than the $3/2$ realistic
value corresponding to manganites~\cite{zang}. Previous calculations~\cite{dago}
have shown that qualitatively $S=1/2$, $3/2$ and $\infty$ (classical spins)
give
very similar phase diagrams in the
absence of Coulomb interactions, and there is no reason to expect that
this situation will change when these interactions are added.
In addition, again due to the large size
of the Hilbert spaces involved in the problem
our analysis must be restricted to
one-dimensional systems. Previous work has shown the existence of many
similarities between the results obtained in all
dimensions~\cite{yunoki} including $1$. Thus, the restriction of
 working on chains should not be considered 
 severe, and the analysis
below is expected to capture the main qualitative aspects of the problem under
investigation.

Note that calculations using $S=1/2$ localized spins on chains
have relevance not only in the context of manganites but
also for recently synthesized one-dimensional materials such as
${\rm Y_{2-x} Ca_x Ba Ni O_5}$ which have two active
electrons per Ni-ion. This compound has been 
studied experimentally~\cite{batlogg} and
theoretically~\cite{nio}, and upon doping interesting properties have been observed
including a metal-insulator transition.

To facilitate the understanding of the main results of this paper
 the phase diagram in one-dimension
 of the Ferromagnetic Kondo model (without Coulombic interactions
and using  localized $S=1/2$ spins) is reproduced in Fig.1 from Ref.~\cite{dago}. 
As explained before, three main regimes were found
using ED and DMRG techniques.
The region labeled FM corresponds to saturated
ferromagnetism (the ground state spin is maximized). 
The IC regime suggests the existence of incommensurate spin
correlations, at least of short-range, and it manifests itself through
the behavior of the spin structure factor $S(k)$ which has a peak that
moves away from $k=\pi$ as holes are added to the ``half-filled'' system.
Finally, the regime labeled 
PS corresponds to  phase separation which was studied 
calculating the density 
$ n $ vs the chemical potential $\mu$, as well as the inverse compressibility,
searching for unstable densities~\cite{dago}.

\begin{figure}[htbp]
\vspace{-0.5cm}
\centerline{\psfig{figure=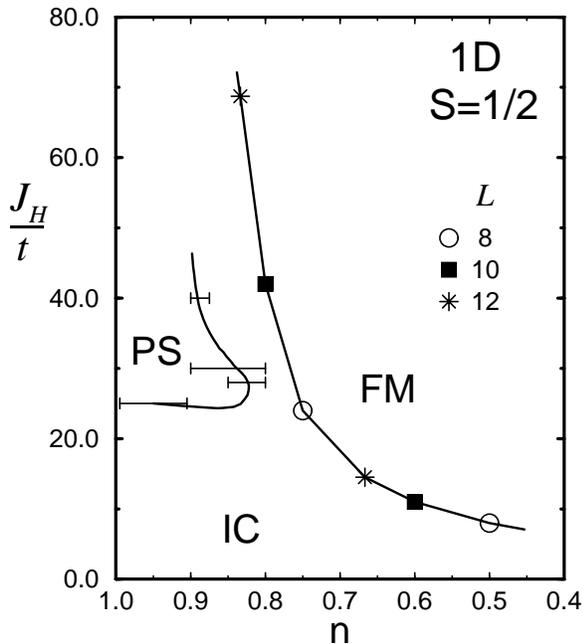,width=8.0cm,angle=-0}}
\vspace{0.2cm}
\caption{
Phase diagram of the Ferromagnetic Kondo model with localized
$S=1/2$ $t_{2g}$-spins. $\rm FM$, $\rm PS$, and $\rm IC$ denote regions with
saturated ferromagnetism, phase separation, and spin-incommensurate 
correlations, respectively. The results for the FM phase were obtained
exactly with the ED technique with up to 12 sites. The rest with
the DMRG method with up to 40 sites. The phase diagram is taken
from Ref.~[6].
}
\end{figure}

The goal of the present manuscript is to analyze how the phase diagram of
Fig.1 changes when Coulomb interactions are added to the one-orbital
 Kondo model.
The organization of the paper is as follows. In section II the
details of the model, technique, and calculated observables are given. In 
Sec. III the phase diagram is constructed when a nonzero on-site repulsion
is considered, but without including
nearest-neighbor Coulomb interactions. In Sec.IV these longer-range
interactions are 
added. The tendency to form charge-ordered states (at least at short
distances) reported in this section is the main result of this paper.
Implications
for experiments are discussed in the conclusions (Sec.V).

\section{Model and Computational Technique}

As explained in the Introduction, 
the model used in the present study is the one-orbital
 Ferromagnetic Kondo model
with localized (spin-1/2) degrees of freedom that mimic the effect of the
$t_{2g}$-electrons. Coulomb interactions in the $e_g$-band are also
incorporated in the model. The  Hamiltonian defined on a chain
with $L$ sites is
$$
H = -t \sum_{{\langle ij \rangle}\sigma} ( c^\dagger_{i \sigma} c_{j
\sigma} + h.c.) - J_H \sum_i { {{\bf S}_{if}}\cdot{{\bf s}_{ic}} } +~~~
$$
$$
~~+ J' \sum_{\langle i j \rangle} { {{\bf S}_{if}}\cdot{{\bf S}_{jf}} }
+ U \sum_i n_{i \uparrow} n_{i \downarrow} + V \sum_{\langle i j
\rangle} n_i n_j.
\eqno(1)
$$
The first term is the electron-transfer between 
Mn-ions. $\langle ij \rangle$ denotes nearest-neighbor sites and
$t$ is the hopping amplitude that will be set to 1 in most of the manuscript.
In the second term,
$J_H>0$ is the ferromagnetic Hund-coupling.
The spin-1/2 operator for the conduction electron is defined as
$
{\bf s}_{ic} = \sum_{\alpha \beta} c^\dagger_{i \alpha} {\bf \sigma}_{\alpha
\beta} c_{i \beta},
$
while
${\bf S}_{if}$ represents 
a localized spin-1/2 at site $i$. A strong Hund-coupling will 
be used throughout the
paper, and its value will be fixed to $J_H = 40$, in units of $t$, unless
otherwise stated.  
$J'$ is the
strength of a direct
Heisenberg coupling between the localized spins. 
This term is needed on
phenomenological grounds since in fully doped manganites (e.g.
when all $\rm La$ has been replaced by $\rm Ca$) 
a finite N\'eel temperature is
experimentally observed. $U$ is
 the strength of the on-site electronic repulsion,
with $n_{i \sigma}$  the number operator at site $i$ with spin $\sigma$.
$V$ regulates the
Coulomb repulsion at a distance of one lattice spacing.
The rest of the notation in Eq.(1) is standard.

The technique used in this paper to analyze ground state properties of
the Hamiltonian Eq.(1) in one-dimension is 
the DMRG method. The finite-system
variation of DMRG was used, working with open boundary 
conditions~\cite{comm101}. All results
were obtained keeping a number of states $m=100$ in the iterations, 
with the exception of densities
$n=1$ and $0$ where
a smaller number of states produced accurate enough
results. With
this value of $m$ a truncation error of order $10^{-6}$ or smaller 
was obtained throughout the results shown in the next sections.

In order to characterize the ground state properties of Eq.(1),
a variety of expectation values have been calculated.
The spin structure-factor defined as
$$
S(k) = {{1}\over{L}}\sum_{j,m} \langle { {{\bf S}_{jf}}\cdot{{\bf
S}_{mf}} } \rangle e^{i(j-m)k},
\eqno(2)
$$
\noindent
and the charge structure-factor 
$$
N(k) = {{1}\over{L}}\sum_{j,m} \langle n_j n_m \rangle e^{i(j-m)k},
\eqno(3)
$$
are among the measured quantities.
In addition, 
the inverse compressibility defined as
$$
1/\kappa = {{N^2}\over{4L}} [ E(N+2,L) + E(N-2,L) - 2E(N,L) ],
\eqno(4)
$$
\noindent was also calculated.
Here $E(N,L)$ is the ground state energy of a chain of $L$
sites with $N$ electrons, and density $n=N/L$.
Finally,
the charge correlation
$C(i) = \langle n_{j} n_{m} \rangle - \langle n_j \rangle \langle n_m
\rangle$, where $i=| j -m|$, was studied in some special cases. 
Here the $\langle \rangle$ notation not only
denotes expectation value in the ground state but it also indicates that
for a given distance $i$ all possible pairs of sites $j,m$ 
of the cluster compatible with $i=|j-m|$ 
have been used. The reason is that open boundary conditions are needed
in the DMRG technique and, thus, the correlations are
different at,  e.g.,  the center and near the chain end.
Such an average procedure uses
information from the whole chain, and in practice
it produces smoothly changing results as
the distance $i$ and the couplings are varied.

\section{ Results at $V=0$}

Let us begin the study for the case without a repulsion at
a distance of one lattice-spacing (i.e. working at
$V=0$). The results obtained here will be later compared with those of
the following section for $V \neq 0$.


\subsection { Zero on-site repulsion $U$}

In Fig.2a, the inverse compressibility $1/\kappa$ is shown as a function
of $n$, with  both the on-site and intersite repulsions $U$
and $V$ equal to zero, 
in order to study the dependence of the results with $J'$. 
A similar analysis was carried out by Yunoki and Moreo~\cite{yunoki2}
but in the classical localized spin limit. This study found that $J'$ is an
important parameter in determining the low-temperature properties of
models for manganites.
Fig.2a shows that
$1/\kappa$ at  $J'=0.04$ is small or even slightly negative
both in the limits of  small and large density, in good
qualitative agreement with Ref.~\cite{yunoki2}.
For larger values of
$J'$, Fig.2a suggests
that the phase separation regime near $n=1$ is lost (i.e. all densities
are stable), while the results close to
$n\sim 0$ are only slightly affected. Then, in order to study the effect
of a nearest-neighbor (NN) 
repulsion $V$ over a phase separated regime, relatively small values
of $J'/t$ must be selected. Note that this is not a problem since
experimental results have
actually shown that $J' \sim 0.05$ in units of $t=0.2eV$ 
is a realistic value for the exchange coupling
between the localized spins~\cite{yunoki2,perring}. 
Then, in the rest of the paper 
$J'=0.05$ will be used, unless otherwise stated.

\begin{figure}[htbp]
\vspace{-0.5cm}
\centerline{\psfig{figure=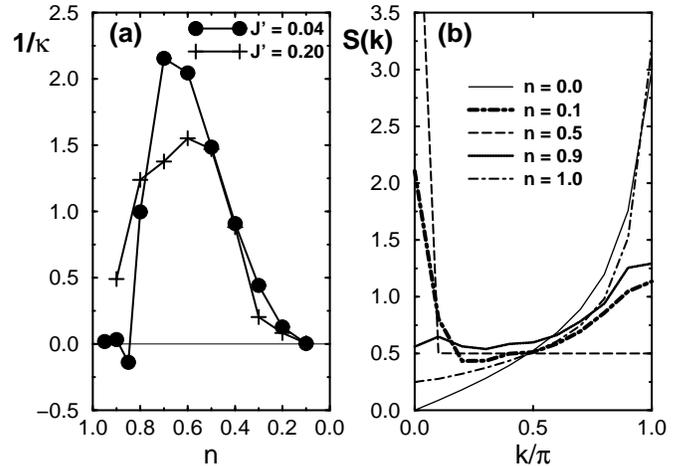,width=8.5cm,angle=-90}}
\vspace{0.2cm}
\caption{
(a) Inverse compressibility $1/\kappa$ versus density $n$ for
the FM Kondo model with $U=V=0$ and  two values of $J'$, as indicated.
The results were obtained using chains with 20 and 40 sites.
For $J'=0.04$ $1/\kappa$ is very small or even negative at $both$
extremes of $n$ close to 1 and 0. At $J'=0.20$ the system is unstable
only at small density; (b) Spin structure factor $S(k)$ for a variety of
densities (indicated), $U=V=0$, $J'=0.04$, and a 20 site chain.
For $n=0.1$ and $0.9$, a
coexistence of ferromagnetic and antiferromagnetic features is
observed. Note that the value 0.5 for $S(k \neq 0)$ in the fully polarized
state at $n=0.5$ originates in the particular structure of the spin
correlation which gives 3/4 on-site and
1/4 for other distances.
}
\end{figure}

Let us now analyze the behavior of the
spin structure-factor $S(k)$ as a function of density. The results are shown
in Fig.2b. As expected, in the limits $n=1$ and $0$ a clear signal for 
strong
antiferromagnetic correlations is observed since $S(k)$ is peaked at
$k=\pi$~\cite{comment99}. 
In the intermediate regime where double-exchange tendencies
 leads to a ferromagnetic
ground state, $S(k)$ is maximized at $k=0$ also as expected based on
previous literature. The most interesting results
in this context arise close to $n=1$ and also $0$.
In this regime, $S(k)$ has
important weight $both$ at $k=0$ and $\pi$ signaling the coexistence
of FM and AF domains as expected in a phase-separated regime. This is
in agreement with the previous work reported in
Ref.~\cite{yunoki}, with the discussion
in the Introduction, and with the inverse
compressibility data shown in Fig.2a. 
Note that the densities $n$ that 
correspond to the phase separation regime
can be studied in detail since the analysis
presented here is in the
$canonical$-ensemble formalism. If a grand-canonical approach would be
used, as in Ref.~\cite{yunoki}, then the regimes
$n \sim 0.1$ and $\sim 0.9$ would not be reachable.

\subsection{Non-zero on-site repulsion $U$}


For the Hund-coupling
used throughout this paper, moderate values of the
 on-site Coulomb interaction
$U$ are not expected to play an important role since double occupancy is
naturally suppressed by a large $J_H$. To illustrate this statement, in Fig.3
the inverse compressibility is shown at $n=0.9$, namely in a regime with
phase separation at $U=V=0$, as discussed in the previous subsection. 
When $U$ is switched on and  increased to a large value in units
of the hopping amplitude
($U=16$), the plots of $1/\kappa$ vs $J'$ 
change only slightly. As a consequence, in
the rest of the paper 
$U$ will be fixed to $16$ to avoid the proliferation of
free parameters. 
The conclusions of our paper are not expected to
change as long as $J_H$ and $U$ are the largest scales in the problem.

\begin{figure}[htbp]
\vspace{-0.5cm}
\centerline{\psfig{figure=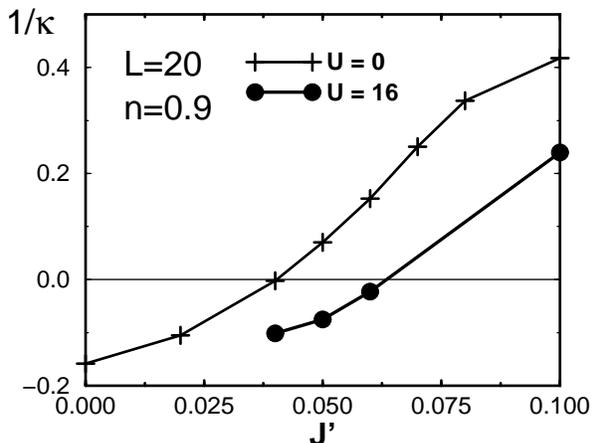,width=8.5cm,angle=-90}}
\vspace{0.2cm}
\caption{
Inverse compressibility $1/\kappa$ versus $J'$ at $n=0.9$, $V=0$,
using a chain with 20 sites, and  two values of $U$ as indicated. The
results illustrate that at large Hund coupling, $U$ does not influence 
much on the ground state properties.
}
\end{figure}

The compressibility at $V=0$ and $U=16$ as a function of $n$ is shown
in Fig.4a. It is clear that both at small and large $n$, this quantity
is either $\sim 0$ or negative signaling the instability of the ground state
towards the formation of two different phases. In between, where the
system is expected to be ferromagnetic, the ground state is stable.
In Fig.4b the spin structure factor $S(k)$ is shown to illustrate the
coexistence of FM and AF features in the ground state of the ``unstable''
regime. This occurs at $n=0.85$ where peaks  at both
$k=0$ and $\pi$ are observed. In the other cases, $n=1$ and $0.75$,
there is only one dominant peak at the antiferromagnetic and
ferromagnetic locations, respectively.

\begin{figure}[htbp]
\vspace{-0.5cm}
\centerline{\psfig{figure=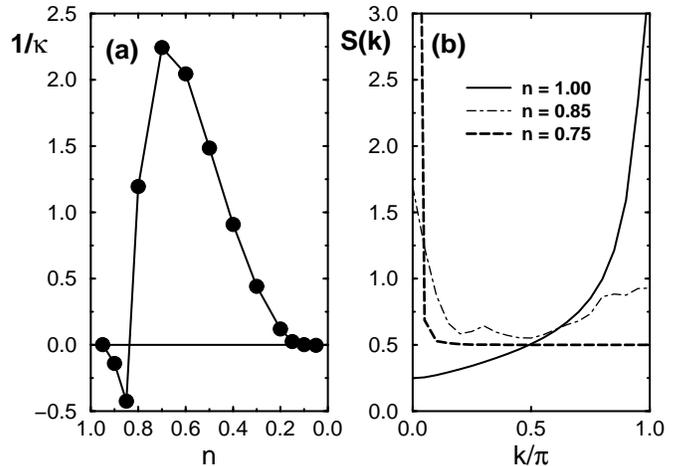,width=8.5cm,angle=-90}}
\vspace{0.2cm}
\caption{
(a) Inverse compressibility $1/\kappa$ vs density $n$, working
at $J_H= 40$, $U=16$, $V=0$, $J'=0.05$, and using 20 and 40 sites.
The results
show that near $n \sim 1$ and $n \sim 0$ phase separation occurs, as 
shown in Fig.2a even including a large on-site repulsion $U$;
(b) Static spin structure factor $S(k)$ vs momentum
$k/\pi$ for the same $J_H, U, V$ couplings
used in (a), $J'=0.05$, 40 sites, and several densities. The
result at $n=0.85$ illustrates the coexistence of FM and AF
features.
}
\end{figure}

To further confirm if at small and large $n$ phase
separation is indeed observed, the on-site densities 
$\langle n_i \rangle$ (expectation value in the ground state of the
local density operator) have also been monitored. 
Note here that the DMRG method works using open boundary conditions and,
as a consequence, the on-site density changes from site to site.  In Fig.5a
results at $n=0.75$, where the compressibility is positive and $S(k)$ peaks
at $k=0$, show that aside from a boundary effect involving about 3 sites
at each chain-end the local density only slightly Friedel oscillates 
around the global density. 
On the other hand, in the presumed to be
phase-separated regime the local density changes substantially as a 
function of the site position 
$i$. Most of the holes are near the boundary and the
local density changes abruptly between two values, which are
the extreme stable densities $\sim 0.75$ and $\sim 1$ as expected.
Also the charge structure factor $N(k)$ (Fig.5b) has the
characteristics corresponding to phase separation, 
namely at $n=0.75$ it behaves as a
non-interacting spinless fermionic system due to the ferromagnetic character of
the ground state, while at $n=0.85$ it     develops 
structure at small wavenumbers
 related with the inhomogeneous distributions of charge.

In short, the results of this subsection have shown that the Kondo model with
 localized $S=1/2$  spins, and 
with the addition of a $J'$-coupling among them, has a qualitative
phase diagram similar to the results presented before
in the classical limit for the $t_{2g}$-spins
in Refs.~\cite{yunoki,yunoki2}. 
This conclusion does not change even if an on-site repulsion $U$ 
of moderate strength is
incorporated in the problem. Then, the present analysis have
allowed us to fix the parameters $J_H,U,J'$ such that the physics
under investigation, focussed on  phase separation and ferromagnetism,
 is realized in the ground state of the model
Eq.(1). Thus, the problem
is now ready for the analysis of the influence of $V$ on the
phase diagram.

\begin{figure}[htbp]
\vspace{-0.5cm}
\centerline{\psfig{figure=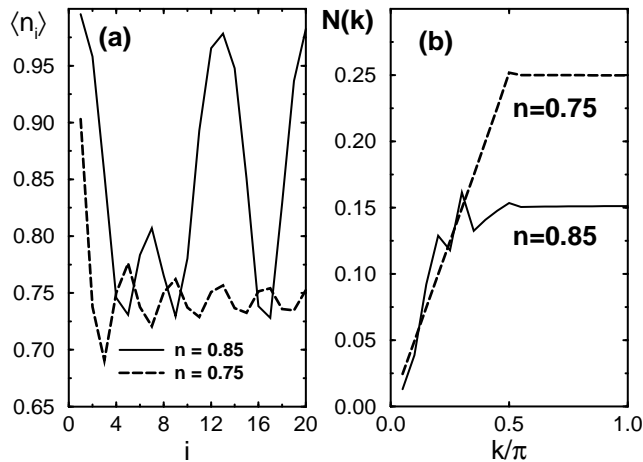,width=8.5cm,angle=-90}}
\vspace{0.2cm}
\caption{
(a) Local density $\langle n_i \rangle$ vs position along the
chain $i$ for the same couplings as in Fig.4a. The chain length is 40.
$1$ is the first site of the chain starting from the left.
$20$ is the center of the chain. The results for $i>20$ are obtained
by reflection. Shown are data for $n=0.75$, representative of the
ferromagnetic region, and $n=0.85$, which belongs to the phase separated
regime; (b) Charge structure factor $N(k)$ at the two densities used in
(a).
}
\end{figure}

\section {Results Including a Repulsion $V$}

\subsection{ Influence of $V$ on the phase separated regime}


The $V$-term will now be switched on at a density such that
phase separation occurs in the ground state. 
In Fig.6a the inverse
compressibility is shown as a function of $V$ using a 40-sites
chain. The results show that the unstable region observed
at $V=0$ in the previous section
now becomes stable when $V > 0.5$. Then, in agreement with the
introductory discussion, phase separation is
severely affected by Coulombic interactions beyond the on-site term.
However, here
it is interesting to observe that vestiges of phase separation
survive even up to large couplings $V$. 
For instance, consider $S(k)$ which is shown in Fig.6b.
At $V=1$, a double-peaked structure is observed, as it occurs at $V=0$,
but now with maxima deviated from $0$ and $\pi$ forming
incommensurate structure. Results at other values of $V >1$ (not shown) 
are very similar to those at $V=1$. Then, the ground state
properties
do not seem to change abruptly with $V$ but smoothly. Even 
ground states that have been stabilized by the Coulomb interaction
 in this regime contain a spin structure-factor with remnants of
FM and AF domains. It is only the macroscopic
accumulation of charge that is penalized by $V$.

\begin{figure}[htbp]
\vspace{-0.5cm}
\centerline{\psfig{figure=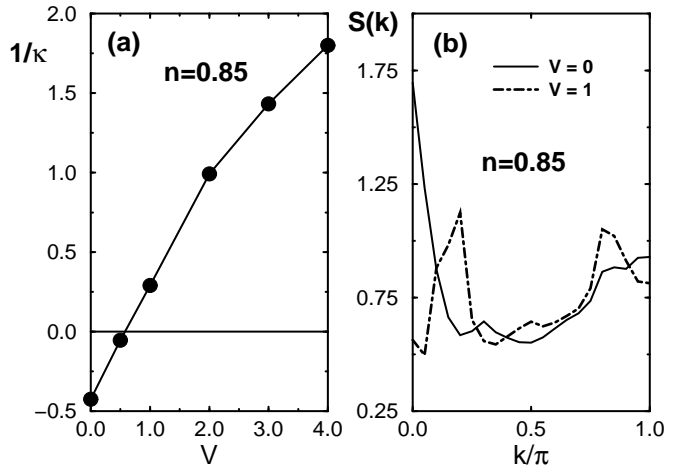,width=8.5cm,angle=-90}}
\vspace{0.2cm}
\caption{
(a) Inverse compressibility $1/\kappa$ vs $V$ working at
$J_H = 40$, $U=16$, $J'=0.05$ and density $n=0.85$ (i.e. in the region of
phase separation for $V=0$). Shown are results using 40 sites.
$1/\kappa$ becomes positive for an intersite Coulomb interaction
 $V \sim 0.5$ making the ground state
stable; (b) $S(k)$ for the parameters used in (a) and two values
of $V$ (actually results for $V>1$ are similar to those of $V=1$).
}
\end{figure}

\begin{figure}[htbp]
\vspace{-0.5cm}
\centerline{\psfig{figure=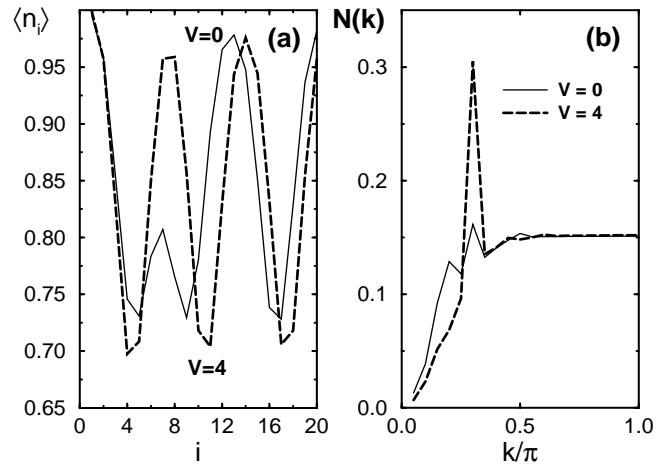,width=8.5cm,angle=-90}}
\vspace{0.2cm}
\caption{
(a) Local density $\langle n_i \rangle$ vs position along the
chain $i$ at $n=0.85$. For the site convention see the caption of Fig.5a. 
The couplings and density are the same as in Fig.6a. The
solid (dashed) line represent results for $V=0$ ($V=4$); (b) $N(k)$ for
the two values of $V$ used in (a). The rest of the couplings and
density are as in (a).
}
\end{figure}

In Fig.7a the local density is shown for the case of 6 holes on a 40-site
chain. The accumulation of charge
near the boundary characteristic of phase separation at $V=0$
is replaced by a fairly clear periodic distribution of charge at $V=4$. 
This occurs not only at $V=4$ but in a wide
range of couplings. The replacement of phase separation by a
state with charge-ordering tendencies
was expected based on the discussion given in the Introduction. 
The density where the ``holes'' are mostly located is close to
$0.7$. There is no clustering of charge at large $V$. From
 Fig.7b it can be observed that the
charge structure-factor $N(k)$ at $V=4$ develops a very sharp peak 
at $k=2\pi n$ due to 
the periodic arrangement of charge in the ground state. 

It is interesting to notice that the positions where the ``holes'' are
in Fig.7a ($V=4$)
are actually made out of four sites instead of one. Then, the holes
are not fully static but they have some mobility, and it is natural that
to enhance this mobility the spins must be aligned. Then, the large FM
regions at small $V$ in the regime of phase separation are now
replaced by periodically distributed small regions resembling 
magneto-polarons. This is among the most important results discussed in
this paper.

\subsection {Influence of $V$ on the ferromagnetic regime}


After analyzing in the previous subsection
the influence of $V$ on the properties
of a phase-separated ground state near $n=1$, let us study what occurs when a
NN-Coulomb interaction  is switched on in 
a fully ferromagnetic ground state. Such a state can be
easily obtained in the present model Eq.(1) simply using a density,
e.g., $n=0.6$ 
where the double-exchange ideas are operative. 
The results for $S(k)$ are shown in Fig.8a. While at
$V=1$ (and smaller values) the peak at $k=0$ characteristic of
ferromagnetism
remains strong, at $V=8$ it has reduced
substantially its intensity and moved slightly from $k=0$.
At $V=12$ a very broad peak is the
only remnant of the ferromagnetic structure at small $V$.

\begin{figure}[htbp]
\vspace{-0.5cm}
\centerline{\psfig{figure=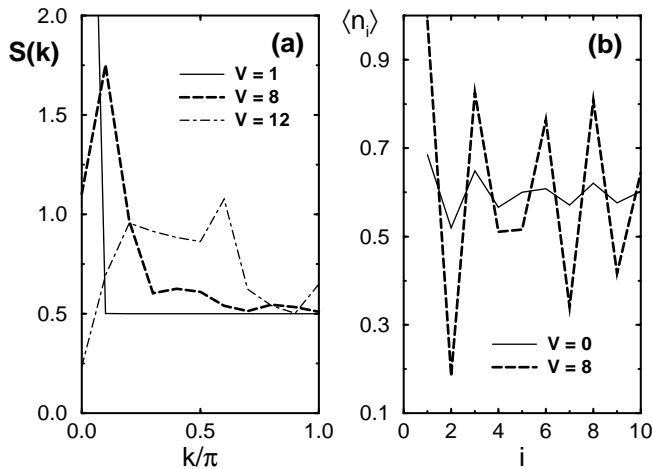,width=8.5cm,angle=-90}}
\vspace{0.2cm}
\caption{
(a) Spin structure factor $S(k)$ working with $J_H=40$, $U=16$,
$J'=0.05$, $n=0.6$, 20 sites, 
and the values of $V$ indicated; (b) Local density
$\langle n_i \rangle$ for the same couplings and density as in (a).
Results for a chain of 20 sites are shown. $i=1$ is the site at the
extreme left of the chain. The results for $i>10$ are obtained by
reflection of the numbers shown.
The results at $V=0$ are almost uniform while at $V=8$ they present
clear large oscillations.
}
\end{figure}

To understand the reason for the complicated behavior of $S(k)$,
 the local density at $n=0.6$ is shown in Fig.8b. At $V=0$ it is 
almost uniform as expected in a ferromagnetic state. However, at
$V=8$ a charge-ordered pattern is observed at $k=2 \pi n$. This is
in agreement with the previously discussed
 results at $n=0.85$ which showed a similar
tendency. However, note that now most minima in the local density involves
only one site, instead of four as in Fig.7a. For this lattice and number
of electrons (12 of them on a 20-site chain), the charges cannot arrange
themselves in a periodic structure,
causing the incommensurate-like peak in $S(k)$ at $V=8$.
The replacement of a metallic FM state by a state with charge-ordering tendencies
is also clear in Fig.9a where $N(k)$ is shown. At $V=8$ a large
peak is observed. Also the real-space density-density correlations
$C(i)$ shown in Fig.9b (for its definition see Sec.II)
 indicate strong effects in the charge sector at short distances: while at $V=0$ these
correlations are not negligible only at distances $i=0$ and $1$, at $V=8$ the
same  correlations
have been enhanced and they are now robust also at distances 2 and 3.

\begin{figure}[htbp]
\vspace{-0.5cm}
\centerline{\psfig{figure=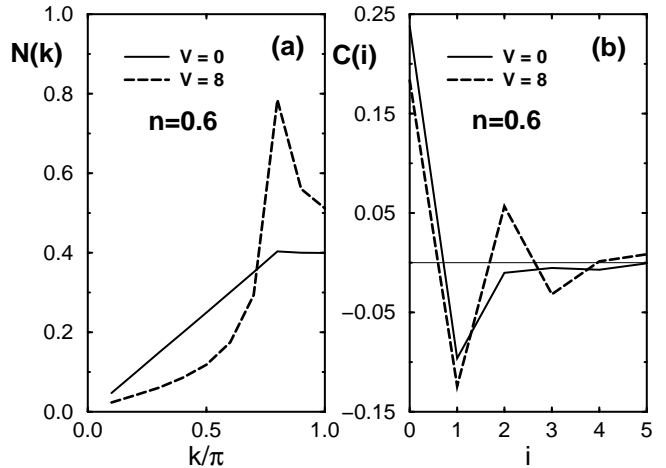,width=8.5cm,angle=-90}}
\vspace{0.2cm}
\caption{
(a) $N(k)$ vs $k/\pi$ for the couplings $J_H,U,J'$
 and density of Fig.8a,
and two values of $V$ (indicated). The chain has 20 sites.
At $V=8$ the large peak suggests strong charge correlations in the
ground state;
(b) Density-density
correlations $C(i)$ (for its definition see text) vs distance $i$,
at the same couplings, chain length, and density used in (a). Enhancement of
charge correlations at $V=8$ is observed.
}
\end{figure}

Working at exactly $n=0.5$ the incommensurate structures observed at
$n=0.6$ should disappear. In Fig.10a the local density is shown for a
20-site cluster and several values of $V$. At $V=0$ the density is 
exactly uniform for symmetry reasons, but as soon as $V$ is switched on
a charge-ordered pattern clearly emerges. At this density a soliton at the
center of the chain appears justifying the reduction in $\langle n_i
\rangle$ towards that center. $N(k)$ in Fig.10b also presents a clear peak
at $k=\pi$ which grows with $V$,
in agreement with the previous discussion. Then, from the analysis of
the densities 
$n=0.6$ and $0.5$ it is concluded that a tendency to charge ordering, at
least at short-distances,
is obtained once an intersite Coulomb interaction $V$ is introduced in a
fully ferromagnetic state. This state 
(which here is referred to as charge-ordered (CO) for simplicity
although the large
distance behavior is difficult to analyze)
occurs $within$ the
ferromagnetic phase i.e. through a calculation of $S(k)$
the spins were found to be fully polarized unless
$V$ reaches larger values than those shown in Fig10a-b. A similar result
was reported in Ref.~\cite{shen} using a path-integral approach.

\begin{figure}[htbp]
\vspace{-0.5cm}
\centerline{\psfig{figure=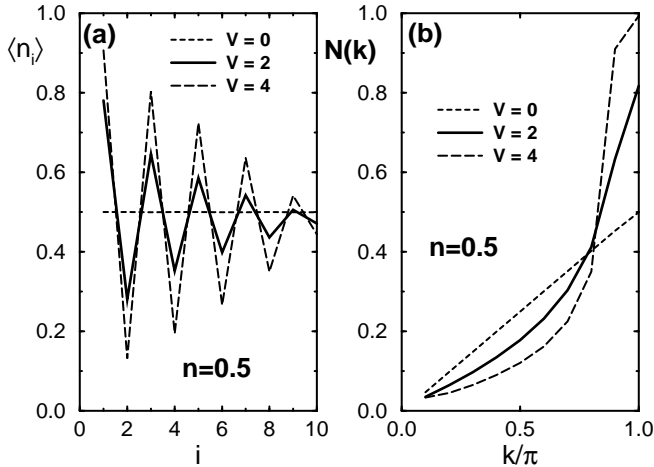,width=8.5cm,angle=-90}}
\vspace{0.2cm}
\caption{
(a) Local density $\langle n_i \rangle$ corresponding to a
20-site chain with $J_H =40$, $U=16$, $J'=0.05$, $n=0.5$, 
and the values
of $V$ indicated in the figure. The development of charge oscillations
as $V$ grows is clear in the figure; (b) $N(k)$ vs $k/\pi$ using
the same couplings, chain length,
 and density as in (a). The development of a large
peak at $k=\pi$ is also compatible with the stabilization of a charge
ordered pattern.
}
\end{figure}

\begin{figure}[htbp]
\vspace{-0.5cm}
\centerline{\psfig{figure=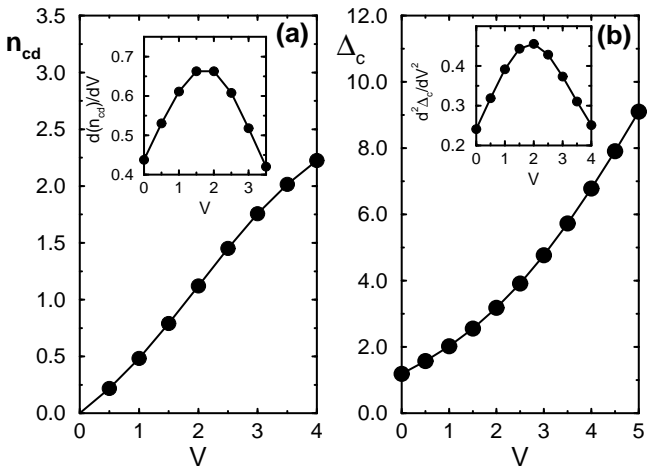,width=8.5cm,angle=-90}}
\vspace{0.2cm}
\caption{
(a) $n_{cd}$ (see text) vs $V$ for the same
$J_H$, $U$, $J'$, chain length, 
and density as in Fig.10a. In the inset the first
derivative is shown; (b) Charge gap $\Delta_c$ as defined in text vs $V$.
The couplings, chain length,
 and density are as in (a). The inset shows the second
derivative. For a discussion see text.
}
\end{figure}

In order to monitor the development of charge correlations in the
ground state the mean value of
$n_{cd} = \sum_{i=odd} ( n_i - n_{i+1} )$ has been studied before
in other models~\cite{mutou}.
Its expectation value should be zero in a metallic state, but nonzero 
in a system with charge order. Results for a chain of 20 sites are
shown in Fig.11a. Although nonzero for all values of $V \neq 0$, this
result suggests only a tendency towards the formation of a
charge-ordered state.
For a proper analysis of the critical $V$ leading to charge ordering 
a careful finite-size
study is needed, beyond the scope (and numerical accuracy) of the
present paper. However,
in order to gain at least some insight from the 20-site cluster, the first
derivative of $n_{cd}$ with respect to $V$ is shown in the inset of
Fig.11a. 
Previous calculations studying variations around the $t-J$ model 
showed~\cite{mutou}
that the critical point is located at the inflection point of the
$n_{cd}$ vs $V$ curve. The calculation shown in Fig.11a suggests  that 
the critical coupling is roughly estimated to be at $V_c \sim 2$, which
is in good agreement with the critical coupling corresponding to a model
of spinless fermions with NN repulsion~\cite{luther}.
The charge gap
$\Delta_c(N,L) = E(N+2,L) + E(N-2,L) -2E(N,L) = {{4L}/{N^2 \kappa}}$
\noindent was also studied (Fig.11b). The second derivative of this quantity
with respect to $V$ is also peaked around $\sim 2$, as in the case of 
$dn_{cd}/dV$. Then, this (rough) analysis suggests that a critical coupling
$V_c \sim 2$ separates two ferromagnetic regions, one charge-disordered
and the other with at least short-distance charge-ordering tendencies. 
However, further work is needed to
establish $V_c$ more accurately.


\section{Phase Separation in the
Small Electronic Density Region}

In the low $e_g$-density region, a nearest-neighbor repulsion is not expected to affect the 
ground state properties substantially since the mean distance between
electrons is large. Coulombic interactions
at distances larger than one lattice spacing would be more
important in this regime (but they are difficult to study with the DMRG
method). To verify that indeed $V$ is not
playing an important role at small $ n $, in
Fig.12a the inverse compressibility is shown for three values
of $V$. While phase separation at large density $n \sim 1$ is affected by
$V$ (as discussed in previous sections) and $\kappa$ changes
substantially at intermediate densities, the results at low
density $ n \leq 0.20$ are almost $V$-independent.

\begin{figure}[htbp]
\vspace{-0.5cm}
\centerline{\psfig{figure=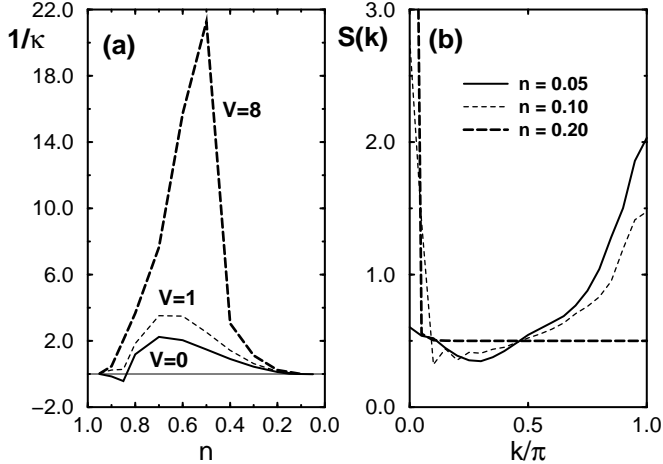,width=8.5cm,angle=-90}}
\vspace{0.2cm}
\caption{
(a) Inverse compressibility $1/\kappa$ vs density $n$, using
$J_H = 40$, $U=16$, $J'=0.05$, and the three values of $V$ indicated. 
The numbers are obtained using chains with 20 and 40 sites.
 The results show that the compressibility is very large at low densities.
$V$ has virtually no influence on this region of the phase diagram; (b)
$S(k)$ for the same couplings $J_H$, $U$, and $J'$ used in (a),
a chain of 40 sites, and using
$V=0$.  The
densities are shown. The results at $n=0.10$ and $0.05$ illustrate
 the coexistence of FM and AF features.
}
\end{figure}

Let us investigate the properties of the phase-separated regime
at small $ n $ and $V=0$. The results for the spin structure
factor are shown in Fig.12b. At $n=0.20$, a clear signal of 
ferromagnetism is observed, with $1/\kappa$ being slightly positive
(stable). However, at $n=0.10$ and $0.05$ a coexistence of weight
both at $k=0$ and $\pi$ appears, signaling the expected 
regime that separates (i) electron-rich spin-ferro 
and (ii) electron-undoped spin-antiferromagnetic states. The presence
of PS can also be inferred from the local density $\langle n_i \rangle$
shown in Fig.13a. At the stable density $n=0.25$, and leaving aside
boundary effects involving about three sites at each end, the
density presents Friedel oscillations around the density $n$ (actually the 
results at this
density could have been obtained directly from those of Fig.5a at
$n=0.75$ since in the fully spin-aligned state low- and high-densities
are exactly related by symmetry).
However,
at $n=0.10$ the 4 electrons present in the $L=40$
system accumulate at
the center, leaving about 10 sites (a quarter of the lattice) 
virtually empty
on each end of the chain. This is the way in which phase separation
seems to manifest itself when clusters with
open boundary conditions are used. The central cluster of electrons 
(which has $n \sim 0.2$, i.e. the lower limit of the densities which are
stable according to Fig.12a) was found to
be spin  ferromagnetic as expected. 
Finally, to further confirm that $V$ does not influence severely on
the low-density regime, in Fig.13b $S(k)$ is shown now at $V=8$
for two densities. If $n=0.20$, the system is ferromagnetic, while
for $n=0.10$, once again a coexistence of FM and AF features is
observed, as in the absence of the intersite Coulomb interaction.

\begin{figure}[htbp]
\vspace{-0.5cm}
\centerline{\psfig{figure=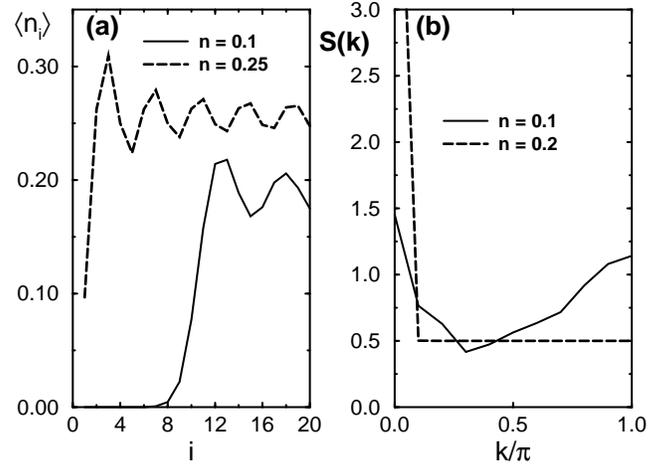,width=8.5cm,angle=-90}}
\vspace{0.2cm}
\caption{
(a) Local density $\langle n_i \rangle$ for the same couplings
$J_H, U, J'$ used in Fig.12a, and $V=0$. The chain used has 40 sites.
Results are shown for $n=0.25$
where the densities are very similar (leaving aside a boundary effect
near $i=1$), and at $n=0.10$ where charge accumulates at the center
due to phase separation tendencies in the ground state; (b) $S(k)$
at $J_H = 40$, $U=16$, $J' = 0.05$, $V=8$, and for the two densities
indicated.  The number of sites is 20.
 The result at $n=0.1$ shows that even including a large $V$,
the coexistence of FM and AF  features persists.
}
\end{figure}

\section{Conclusions}

In this paper the influence of a nearest-neighbor Coulomb repulsion $V$
between $e_g$-electrons was studied using the Ferromagnetic Kondo
model. The main goal of the paper was to analyze the evolution with $V$ of the
ground states in the regimes of (i) the recently
computationally discovered phase-separation~\cite{yunoki,dago,yunoki2} and 
(ii) with ferromagnetism induced by the double-exchange mechanism. Spin and charge
correlations were studied in detail.
The computational work using DMRG was made possible by
restricting the spatial dimension to 1, the localized spin value to 1/2,
and using only one-orbital per site. These limitations prevent us from 
making detailed quantitative statements about the effect of Coulomb
interactions on real manganite compounds. 
Actually it is unrealistic to expect
that accurate numerical work will be possible in the near
future for large enough clusters in dimensions larger than one,
 considering the Coulombic interactions included here.
However, several qualitative features have
emerged that seem robust enough to survive an increase in
dimensionality.

The overall phase diagram in the $V$-$n$ plane found in this study is
presented in Fig.14. In the limit $V=0$, two regimes of phase separation
were observed near $n=1$ (PS1) and $0$ (PS2) (in excellent agreement
with Ref.~\cite{yunoki2}). In
between, a robust ferromagnetic phase was observed with no indications of
strong charge ordering tendencies. However, when the Coulomb repulsion
$V$ is included in the calculation the PS1 regime
rapidly becomes unstable due to the expected energy penalization caused
by the charge difference between the two competing
phases. At these densities, the 
$V$-term induces a regular arrangement of charge which resembles a polaron
lattice. Each hole is spread over four lattice sites in the regime of
parameters investigated in Fig.7a.
This picture gives support to the intuitive notion that
visualizes a phase separated state with extended Coulomb interactions included as a
collection of small ``islands'' of one phase embedded into the other.
This results is in good agreement with a large body of experimental work
in manganites~\cite{hennion,allodi,lynn,deteresa,cox,yamada,bao,otroexp,perring}.
Vestiges of the phase separated regime are observed in the
spin structure factor which has weak
incommensurate-like peaks both near
$k=\pi$ and $0$. In the other extreme of low electronic density, PS2 is not affected by $V$ since
the mean distance between carriers is large at low density. Then, here phase
separation persists up to large values of $V$. Presumably longer-range Coulombic
terms are needed to melt this regime, and induce the charge-ordered
pattern found in experiments for manganites.

Another of the main results of the paper is the stabilization
by NN-Coulomb interactions of a 
ferromagnetic state with charge periodically distributed  at least at
short-distances. This state is expected to be an insulator but a
calculation of the Drude weight is needed to confirm this conjecture.  
This occurs in the vicinity of $n=0.5$ (Fig.14). The
electrons gain kinetic energy by keeping the spin background
fully polarized and they avoid the Coulomb repulsion by forming
a charge pattern which is approximately periodic.
For $n \neq 0.5$, the charge structure-factor
peaks away from $k=\pi$, signaling a (natural) tendency to form 
incommensurate charge arrangements, but its strength is weaker than at
$n=0.5$. Then, the present calculation suggests that the ferromagnetic
phase of the manganites may coexist with charge ordering tendencies
which are maximized at $n=0.5$. Likely these tendencies are more dynamic
than static. Note that recent experimental results by C. H. Chen and
S-W. Cheong (Ref.~\cite{chen}) have reported the existence of weakly
incommensurate charge ordering in $\rm La_{0.5} Ca_{0.5} Mn O_3$
using electron diffraction techniques. Our results suggest that this
phenomenon may originate on the influence of NN-Coulomb interactions
on double-exchange induced ferromagnetic phases of the manganites.

\begin{figure}[htbp]
\vspace{-0.5cm}
\centerline{\psfig{figure=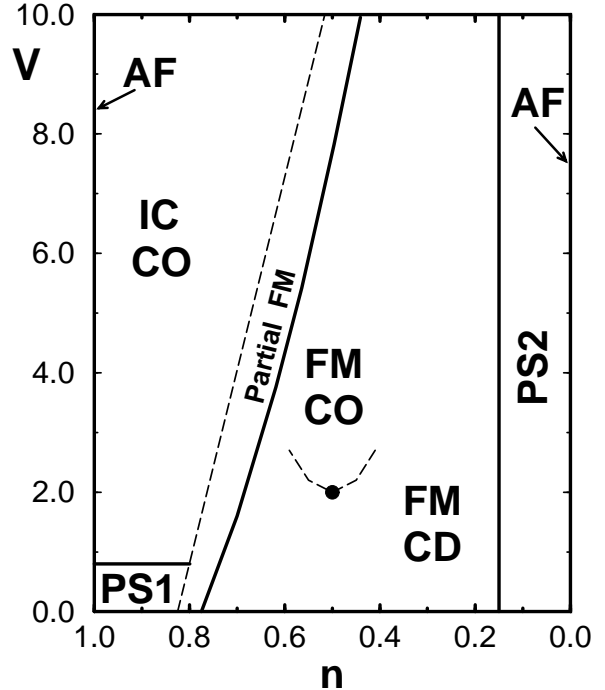,width=8.5cm,angle=-0}}
\vspace{0.2cm}
\caption{
Phase diagram of the Ferromagnetic Kondo model with on-site and
intersite Coulombic interactions, using $S=1/2$ localized spins, $J_H = 40$, 
$J' =0.05$, $t=1$, and
in one-dimension. The results are semi-quantitative since only a finite
number of densities can be achieved with clusters of 20 and 40 sites.
Thus, small scale oscillations have been smeared out in constructing
this figure. $\rm PS1$ denotes the
regime of phase separation at  $n \sim 1$, while $\rm PS2$ indicates a similar
regime but now at $n \sim 0$.  Both extremes $n=1$ and $n=0$ have strong
antiferromagnetic (AF) correlations as indicated. $\rm FM$ denotes a regime where the
spin is fully saturated (the total spin is maximum). $Partial$ $\rm FM$ is a
regime where the total spin is between the maximum compatible with
the cluster and density used, and the minimum (zero). $\rm IC$ indicates
a tendency to form spin-incommensurate structures (for a discussion see
text). $\rm CD$ means ``charge-disordered''.
Reciprocally, $\rm CO$ means ``charge-ordered'', which in this case only means
that in the ground state strong short-distance tendencies to charge
ordering have been identified (the large distance behavior remains to be
analyzed). 
The point and dashed
line at $n=0.5$ and $V=2$ should only be considered as a
tentative rough 
boundary between the $\rm CD$ and $\rm CO$ regimens. It 
represents the evidence discussed in Fig.11 of a singularity in some
observables separating the $\rm CD$ and $\rm CO$ regions within the fully saturated
ferromagnetic state.
}
\end{figure}


\section{acknowledgments}

A. L. M. acknowledges the financial support
of the Conselho Nacional de Desenvolvimento Cient\'\i fico
e Tecnol\'ogico (CNPq-Brazil), as well as partial support from 
the NHMFL
In-House Research Program, supported under grant DMR-9527035.
S. Y. and E. D. are supported by the 
NSF grant DMR-9520776.

\medskip

\vfil


\begin{references}
\bibitem{comm101} The ground state was always assumed to be even under
reflections with respect to the center of the chain.
\bibitem{comment99} 
Note that strictly speaking at large $J_H$ and $n=1$ the system
may resemble a spin-1 chain with a finite spin-gap. However, even in
this state the antiferromagnetic correlations are very robust. For the
purposes of the present paper only the dominant character of the spin and
charge correlations is important, and not their asymptotic behavior with
distance.
\end{references}
\end{document}